\documentclass[fleqn,usenatbib]{mnras}

\usepackage{lmodern}

\usepackage[T1]{fontenc}
\usepackage[dvipsnames]{xcolor}

\DeclareRobustCommand{\VAN}[3]{#2}
\let\VANthebibliography\thebibliography
\def\thebibliography{\DeclareRobustCommand{\VAN}[3]{##3}\VANthebibliography}

\usepackage{graphicx}	
\usepackage{amsmath}	
\usepackage{orcidlink}
\usepackage{subcaption}

\definecolor{bd}{HTML}{8A0303}

\title[Microlensing of Microlensing]{Microlensing of Microlensing: Effects of Random Stars on the Double-Source-Plane Gravitational Lens}

\author[N. Salama et al.]{
Nada Salama\orcidlink{0000-0000-0000-0000},$^{1}$\thanks{E-mail: nada.salama@sydney.edu.au (NS)}
Daniel J. Ballard\orcidlink{0009-0003-3198-7151},$^{1}$
Huimin Qu\orcidlink{0009-0006-0299-0265},$^{1}$
Geraint F. Lewis\orcidlink{0000-0003-3081-9319}$^{1}$
\&
Karl Glazebrook\orcidlink{0000-0002-3254-9044}$^{2,3}$
\\
$^{1}$Sydney Institute for Astronomy, School of Physics A28, The University of Sydney, NSW 2006, Australia\\
$^{2}$Centre for Astrophysics and Supercomputing, Swinburne University of Technology, PO Box 218, Hawthorn, VIC 3122, Australia\\
$^{3}$The ARC Centre of Excellence for All Sky Astrophysics in 3 Dimensions (ASTRO 3D), Australia\\
}

\date{Accepted XXX. Received YYY; in original form ZZZ}

\pubyear{\the\year{}}

\begin{document}
\label{firstpage}
\pagerange{\pageref{firstpage}--\pageref{lastpage}}
\maketitle

\begin{abstract}
Microlensing, the influence of stars within a galactic gravitational lens, has emerged as a powerful probe of compact mass and, through differential magnification, sub-parsec scale sources at cosmological distances. The recent discovery of a double-source-plane gravitational lens system in which the most distant source is a quasar offers the prospect of \textit{compound} microlensing, in which quasar light rays are influenced by compact masses within the two foreground lensing galaxies. Here, we present the first numerical simulations of this ``microlensing of microlensing''. We consider the recently discovered ``Einstein zig-zag'' lens, J1721+8842, as a fiducial case, and construct microlensing magnification maps for each of the six quasar images in this system. Due to the secondary microlensing effects of the myriad of initial microimages, the resulting maps contain more complex caustic features than seen in the case of single plane microlensing. This is reflected in the expected lightcurves seen for each of the images. 
Given the cosmological promise of time-delay lenses for measuring the Hubble constant, combined with more powerful constraints on the lensing potential in double-source-plane lenses, examining the statistical characteristics of compound microlensing is timely to maximise the cosmological output of double-source-plane time-delay lenses.

\end{abstract}

\begin{keywords}
    {cosmology: theory, 
    quasars: general,
    gravitational lensing: micro,
    gravitational lensing: strong}
\end{keywords}



\section{Introduction}
\label{sec:intro}

Over the past few decades, strong gravitational lensing, where a source is multiply imaged and strongly magnified, has become a cornerstone in our study of galaxy evolution and cosmology. Observations of a statistically significant sample of strong gravitational lens systems in the Euclid and Rubin-LSST era \citep{collet_population_2015} are expected to further enhance the position of strong lensing as a competitive cosmological probe.

Rubin-LSST is set to discover $\mathcal{O}(10^{3})$ gravitationally lensed quasars over its lifetime \citep{oguri_gravitationally_2010, yue_mock_2022}. 
Early Euclid data is showing that approximately one in 100 discoverable galaxy-scale strong gravitational lenses have \textit{two} background sources at different redshifts \citep{euclid_collaboration_euclid_2025}; these are double-source-plane lenses (DSPLs). More generally, \textit{multi}-source-plane lenses, or \textit{compound lenses} (referring to a lensed object becoming recursively lensed by a second deflector), are powerful probes of the dark energy equation of state parameter, $w$, and the matter density parameter, $\Omega_{m}$, through the angular diameter distance ratio between lenses and sources along the line-of-sight \citep{golse_constraining_2002, collett_cosmological_2014, sahu_cosmography_2025, bowden_constraining_2025, urcelay_carousel_2026}. 

In a fraction of DSPLs, one of the sources (or even rarer still, both) will be quasars, and hence subject to the influence of gravitational \textit{micro}lensing. This is due to the gravitational lensing influence from compact objects, such as stars and black holes, as they move in the foreground galaxy \citep{young_q0957561_1981, paczynski_gravitational_1986, wambsganss_microlensing_1990}. The time-dependent micromagnifications are sensitive to both the compact mass distribution in the lens and the physical scale of the source. Hence, quasar microlensing has become an established probe of subparsec scale structure in the distant universe \citep[see e.g.][]{yonehara_x-ray_1998, mortonson_size_2005, schnieder_gravitational_2006, mediavilla_structure_2011, rojas_strong_2014}.

Here, we examine, for the first time, the role of microlensing of background quasars in DSPLs and the resultant brightness variations induced. In Section~\ref{sec:lensing} we outline the background to gravitational macro- and microlensing, including the mathematical formalism of DSPLs and the resultant compound microlensing effect. In Section~\ref{sec:method}, we present the adopted numerical approach to produce the effects of compound microlensing from a macromodel of J1721+8842. Our results are presented in Section~\ref{sec:results}, and we discuss and conclude in Section~\ref{sec:discussion}.

\begin{figure*}
    \centering
    \includegraphics[width=0.95\linewidth]{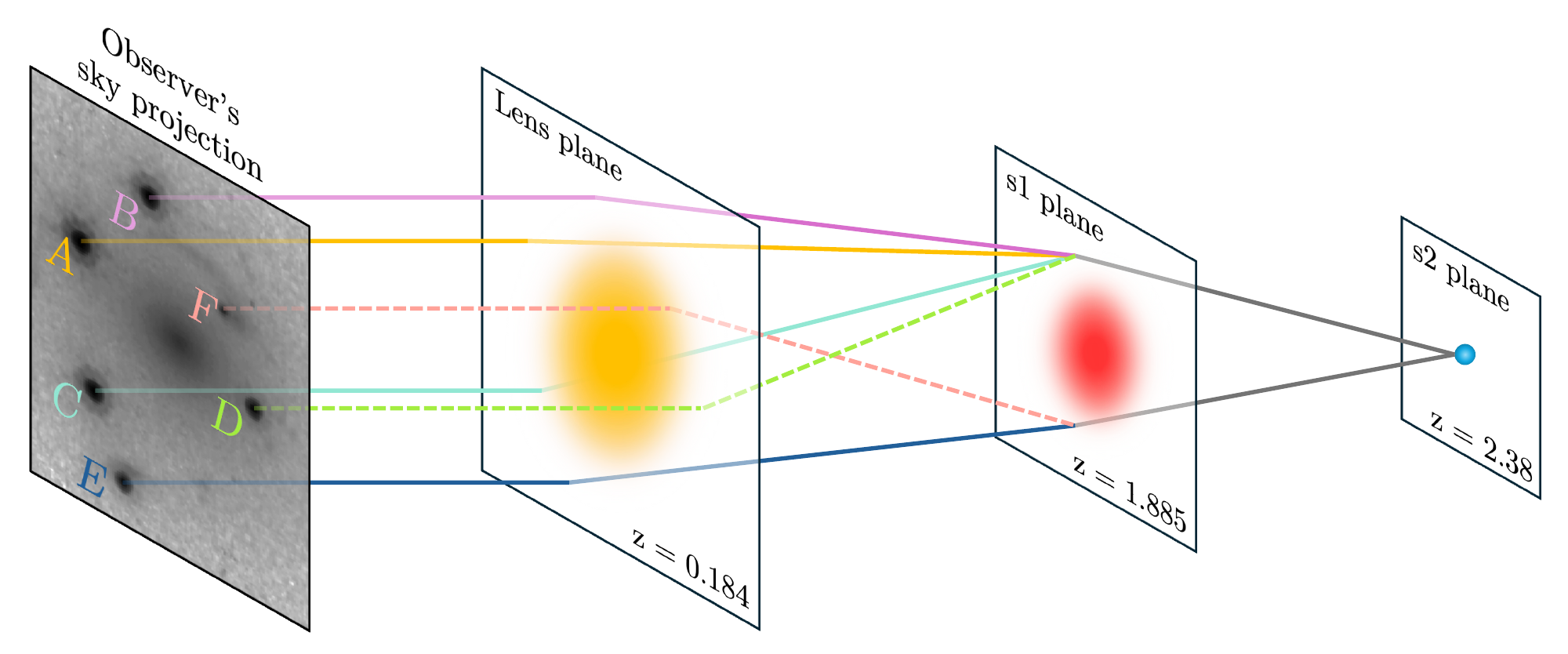}
    \caption{Illustration of the fiducial DSPL configuration employed in this study, the Einstein zig-zag lens J1721+8842. Light rays from source 2 (s2, a quasar at $z=2.38$) are deflected by source 1 (s1, a galaxy at $z=1.885$) and the foreground lens at $z=0.184$, producing six lensed images labelled $A$--$F$. The observer's sky projection shows the resultant image. The dashed rays denoting the trajectory of images $D$ and $F$ to the s2 plane, show the ``zig-zag'' procedure in action, where the images switch places between the s1 plane and the foreground lens plane. We discuss our approach to modelling of this system in Appendix~\ref{app:macromodel}.}
    \label{fig:ray_tracing_illustration}
\end{figure*}

\begin{figure}
    \centering
    \includegraphics[width=0.9\linewidth]{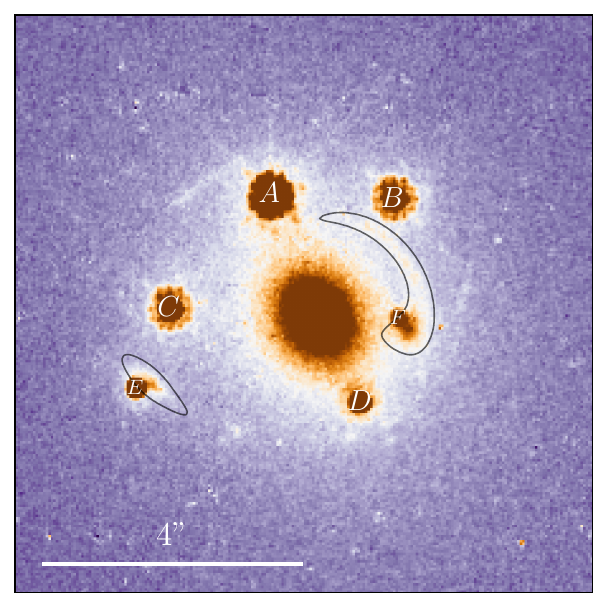}
    \caption{The F814W HST data of J1721 used to fit a macrolensing model, with quasar images $A$--$F$ labelled as in \citet{dux_j17218842_2025} The black contours highlight the lensed images of s1, plotted for illustrative purposes only.}
    \label{fig:J1721_data}
\end{figure}

\section{Compound Microlensing}
\label{sec:lensing}

\subsection{Background}
\label{sec:background}

The gravitational lensing of time-varying sources has long been recognised as a powerful and largely independent probe of cosmology as the lensing-induced time delay depends on the inverse of $H_{0}$ \citep{refsdal_possibility_1964} and intensive campaigns are now meeting this promise \citep{treu_time_2016}. In \citet{shajib_STRIDES_2020} the smallest error bar on $H_{0}$ from a single lensing system so far is $3.9\%$, owing to extra background sources offering bonus constraints on its lens model, and the TDCOSMO collaboration has achieved $2.4\%$ precision on $H_{0}$ using six lensed quasars \citep{wong_h0licow_2020}.

However, the requirements that sources are time-varying mean that, for any measurably short variations, sources must be astrophysically small, such as supernovae or the cores of quasars. This small scale means that these sources are susceptible to the gravitational influence of individual stars in the lensing galaxies. This \textit{microlensing} can induce time-varying magnification of these sources \citep{chang_flux_1979, paczynski_gravitational_1993, tie_microlensing_2018}, complicating the measurement of the sought-after time delays. Microlensing appears to be ubiquitous, resulting in pronounced flux anomalies in gravitational lens systems \citep{schechter_quasar_2002, pooley_x-ray_2012,mediavilla_quasar_2024}. The uncertainty introduced on the inferred time delays by microlensing is negligible in the regime of $\sim10\%$ precision $H_0$ measurements. However, the current goal in the strong lensing community is to reach $\sim1\%$, using at least $\sim40$ systems to do so \citep{treu_strong_2022}. With measurable time delays in galaxy-scale lenses of $\mathcal{O}(\text{days})$, and worst-case microlensing yielding microlensing-induced time delay of $\mathcal{O}(\text{hours})$ \citep{bonvin_cosmograil_2018}, time delay cosmography will eventually need to take the microlensing signal into account.

Cosmological studies with gravitational lenses have focused on single-source-plane systems. However, with the expansion of deep surveys, the discovery of DSPLs with time-varying sources was inevitable.
Recently, a sextuply-imaged quasar J1721+8842 (henceforth J1721) was reported \citep{dux_j17218842_2025} and modelled \citep{schmidt_TDCOSMO_2025}; this is a DSPL with a quasar on the higher-redshift source plane. The foreground lens, a massive elliptical galaxy, is at redshift $z_{\text{lens}}=0.18$; the first source, another early-type galaxy, is at $z_{\text{s1}}=1.89$; the second source, the quasar, is at $z_{\text{s2}}=2.38$. Uniquely, when compared to the present sample of compound lenses, s1 appears to have a particularly substantial lensing effect on s2. This yields multiple imaging of s2 \textit{before} its light reaches the foreground lens and gets multiply-imaged \textit{again}, producing a ``zig-zag'' effect in the trajectories of some of the light rays around s1 and the foreground lens. This behaviour had been theorised and mathematically formalised \citep{collett_compound_2016}, but never observed in a real system until now. At present, there are $\mathcal{O}(10)$ confirmed or candidate DSPLs \citep{euclid_collaboration_euclid_2025, AGEL_barone_2025}. Notable independent DSPL discoveries include SDSS J0946+1006 \citep[the “Jackpot Lens”;][]{gavazzi_sloan_2008}, SDSS J0100+1818 \citep{bolamperti_reconstructing_2023}, HSC J142449-005322 \citep[“Eye of Horus”; ][]{wong_imaging_2017}, DES J0408-5354 \citep{shajib_STRIDES_2020}, J0920+4521 \citep[candidate;][]{lemon_gravitationally_2023}, as well as J1721+8842 \citep[first “Einstein zig-zag”;][]{dux_j17218842_2025}.

The discovery of DSPL systems should aid in the breaking of degeneracies inherent in single-source-plane systems, such as source position transformations that leave image positions and flux ratios unchanged \citep{schneider_source_2014}. 
With a second Einstein ring providing enhanced constraints on the mass distribution of the foreground deflector, and $H_{0}$-invariant angular diameter distance ratios between every plane held fixed, the source position transform is suppressed \citep{ballard_gravitational_2024}. 
Coupled with the measurement of time delays between images, DSPLs therefore may leverage this to reduce the error budget contribution of the lensing potential on measurements of $H_{0}$.

The recursively deflective nature of a compound lens with a background quasar offers a new channel for microlensing, where a collection of microlensed images from one lens plane's mass are themselves microlensed by another: microlensing of microlensing, henceforth \textit{compound microlensing}. The significant lensing effect of deflectors at different redshifts makes J1721 a particularly good testbed for studying this effect on magnification maps and quasar lightcurves. Unlike single-plane microlensing, which produces folds and cusps arranged into relatively simple caustic networks, \citet{petters_caustics_1995} predict that the compound microlensing magnification maps should generate more intricate and denser caustic morphologies, which qualitatively resemble higher-order caustic metamorphoses \citep[see e.g.][]{petters_singularity_2001}. We therefore expect lightcurves in a DSPL to similarly contain more frequent and stronger peaks. Statistical classification of these systems will be especially important if DSPLs are to reach their full cosmological potential.

\subsection{DSPL Theory}

For a comprehensive introduction to gravitational lensing, including the mathematics of multi-plane lensing, the reader is directed to
\citet{schneider_gravitational_1992}. Here, we focus on the key mathematics for this current study. For this study, we consider the gravitational lensing geometry of J1721 as a fiducial representation of DSPL; this is illustrated in  Figure \ref{fig:ray_tracing_illustration}. We discuss our fiducial macromodel for this system in Appendix~\ref{app:macromodel}.

\subsubsection{Macrolensing}
\label{subsec:macro}

We follow the geometry of a DSPL as illustrated in Figure~\ref{fig:ray_tracing_illustration}. Here, the most distant source (s2) undergoes macro- and microlensing events on two distinct redshift planes: the foreground-most lens plane, and the intermediate source plane (s1), whereas s1 undergoes lensing from the foreground-most lens plane only. The convergence, $\kappa_{l}$, of a lens at redshift $z_{l}$ with a background source at redshift $z_{s}$ is written in terms of the 2D-projected mass density on the lens plane, $\Sigma_{l}$:
\begin{equation}
    \kappa_{l}=\frac{\Sigma_{l}}{\Sigma_{\text{cr}}(z_{l}, z_{s})},
    \label{eqn:kappa}
\end{equation}
where
\begin{equation}
    \Sigma_{\text{cr}}(z_{l}, z_{s})=\frac{c^2}{4\pi G}\frac{D(0, z_{s})}{D(0, z_{l})D(z_{l}, z_{s})}
    \label{eqn:critdens}
\end{equation}
is the critical surface mass density, or the threshold surface mass density above which a lens can produce multiple images. $D(z_{a}, z_{b})$ are the relative angular diameter distances between the redshifts $z_{a}$ and $z_{b}$. For a flat $\Lambda$CDM cosmology, these are computed as
\begin{equation}
    D(z_{a}, z_{b})=\frac{c}{H_{0}}\frac{1}{1+z_{b}}\int_{z_{a}}^{z_{b}}\frac{dz}{\sqrt{\Omega_{m}(1+z)^{3}+(1-\Omega_{m})}}.
    \label{eqn:angular_diameter_distance}
\end{equation}
The projected surface mass density, $\Sigma_{l}$, is an intrinsic property of an object at $z_{l}$, invariant to source redshift. It therefore follows directly from Equation \ref{eqn:kappa} that the quantity $\kappa_{l}\Sigma_\text{cr}(z_{l}, z_{s})$ is conserved as $z_{s}$ changes.

As a consequence of this, a lens with two background sources at different redshifts yields two values for $\kappa_{l}$. By default, our chosen convention is that $\kappa_{l}$ refers to the convergence acting on the source most immediately behind. Therefore, the foreground lens has convergence $\kappa_{\rm lens}=\Sigma_{\rm lens}/\Sigma_{\rm cr}(z_{\rm lens}, z_{\rm s1})$ and the mass on the s1 plane has convergence $\kappa_{\rm s1}=\Sigma_{\rm s1}/\Sigma_{\rm cr}(z_{\rm s1}, z_{\rm s2})$. To compute the convergence of the lens acting on s2, we therefore rescale $\kappa_{\rm lens}$ by a factor, $\eta$, which is the ratio of $\kappa_{\rm lens}$ for the two different background source redshifts:
\begin{equation}
\label{eq:eta}
    \eta=\frac{\Sigma_{cr}(z_{\text{lens}}, z_{\text{s1}})}{\Sigma_{cr}(z_{\text{lens}}, z_{\text{s2}})}=\frac{D(0, z_{\text{s1}})D(z_{\text{lens}}, z_{\text{s2}})}{D(0, z_{\text{s2}})D(z_{\text{lens}}, z_{\text{s1}})}. 
\end{equation}

The double-source-plane lens equations are written as:
\footnote{Note that an alternative scaling parameter, $\beta\equiv\eta^{-1}$, is utilised in e.g. \citet{schneider_gravitational_1992}. Then, the multi-plane lens equations in Equation \ref{eq:lens_eq_dspl} become $\boldsymbol\theta_{\rm s2}=\boldsymbol\theta_{\rm lens} - \boldsymbol\alpha_{\rm lens}(\boldsymbol\theta_{\rm lens})-\boldsymbol\alpha_{\rm s1}(\boldsymbol\theta_{\rm s1})$ and $\boldsymbol\theta_{\rm s1}=\boldsymbol\theta_{\rm lens} - \beta\boldsymbol\alpha_{\rm lens}(\boldsymbol\theta_{\rm lens})$. Whereas we use $\eta$ in this work to re-scale $\alpha_{\rm lens}$ defined using $\rm s1$ as the source, the $\beta$ quantity \textit{down}-scales $\alpha_{\rm lens}$ defined with $\rm s2$ as the source. Therefore, $\beta\times\alpha_{\rm lens}$ and our $\eta\times\alpha_{\rm lens}$ both equate to the same $\boldsymbol{\hat{\alpha}}_{\rm lens}$.}
\begin{equation}
\label{eq:lens_eq_dspl}
{\begin{split}
    \boldsymbol\theta_{\text{s2}}&=\boldsymbol\theta_{\text{lens}} - \eta\boldsymbol\alpha_{\text{lens}}(\boldsymbol\theta_{\text{lens}})-\boldsymbol\alpha_{\text{s1}}(\boldsymbol\theta_{\text{s1}}),\\
    \boldsymbol\theta_{\text{s1}}&=\boldsymbol\theta_{\text{lens}} - \boldsymbol\alpha_{\text{lens}}(\boldsymbol\theta_{\text{lens}})
\end{split}}
\end{equation}
where $\boldsymbol\theta_{i}$ are angular positions on plane $i$, and $\boldsymbol\alpha_i$ are the reduced deflection angles due to the mass on plane $i$. As with our chosen scaling convention for $\kappa$, $\boldsymbol\alpha_i$ are defined relative to the distance-scaled deflection angles, $\boldsymbol{\hat{\alpha}}_{i}$, using the closest source redshift behind the deflector, such that $\boldsymbol\alpha_{\rm lens}=[D(z_{\rm lens}, z_{\rm s1})/D(z_{\rm s1})]\boldsymbol{\hat{\alpha}}_{\rm lens}$ and $\boldsymbol\alpha_{\rm s1}=[D(z_{\rm s1}, z_{s2})/D(z_{\rm s2})]\boldsymbol{\hat{\alpha}}_{\rm s1}$.

Partial derivatives of Equations \ref{eq:lens_eq_dspl} can be used to construct a $2\times2$ Jacobian matrix, $\mathcal{A}_{ls}$, describing the mapping from a position on a lens plane at $z_{l}$, $\theta_{l}$, to a position on a source plane at $z_{s}$, $\theta_{s}$. Using 2D Cartesian coordinates, $\boldsymbol\theta_{i} = (x_{i}, y_{i})$, we can write this Jacobian as:
\begin{equation}
\label{eq:jacobian}
    \mathcal{A}_{ls}=\begin{pmatrix}
        \frac{\partial x_s}{\partial x_l} & \frac{\partial x_s}{\partial y_l} \\
        \frac{\partial y_s}{\partial x_l} & \frac{\partial y_s}{\partial y_l}
    \end{pmatrix}
\end{equation}
The magnification, $\mu$, of a source from plane $s$ after being deflected by a lens on plane $l$ is defined as the inverse of the determinant of the Jacobian describing this coordinate transform:
\begin{equation}
\label{eq:magnification_from_jacobian}
    \mu=\frac{1}{\text{det}\mathcal{A}_{ls}}.
\end{equation}
The ray shooting equations in Equation~\ref{eq:lens_eq_dspl} can be used to fit a macro-model of $\alpha_{\rm lens}$ and $\alpha_{\rm s1}$ for a background quasar lensed by a DSPL. The Jacobian matrices in Equation~\ref{eq:jacobian} are computed by automatic differentiation of these equations using \texttt{JAX}\footnote{\url{https://docs.JAX.dev/en/latest/}}, which provides exact numerical derivatives of the lens mapping at each image position. From these, we can calculate via Equation~\ref{eq:magnification_from_jacobian} the ``effective'' magnification observed at each quasar image position, $\mu_{\rm eff} = \det(\mathcal{A}_{\rm lens, s2})^{-1}$. Crucially, to go on to simulate the microlensing on each plane, we also compute the single-plane magnifications for the lens acting on s1, $\mu_{\rm lens} = \det(\mathcal{A}_{\rm lens, s1})^{-1}$, and of s1 acting on s2, $\mu_{\rm s1} = \det(\mathcal{A}_{\rm s1, s2})^{-1}$.

\begin{figure*}
    \centering
    \includegraphics[width=1.0\linewidth]{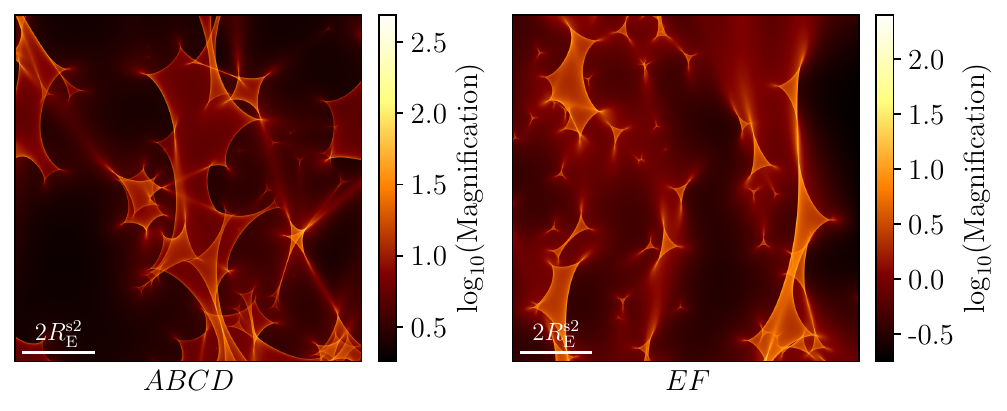}
    \caption{The s1-plane magnification maps at images $ABCD$ (left) and images $EF$ (right). Parameters used to construct these maps are taken from Table~\ref{tab:svi_median_kappa_gamma_s1}}\label{fig:s1maps} 
\end{figure*}

\subsubsection{Microlensing}
\label{subsec:micro}

The macromodel sets the overall image geometry and magnifications. At each macroimage position on a given plane, the total surface density $\kappa$ is comprised of a smooth mass component $\kappa_s$ and a point mass component $\kappa_*$, associated with individual stars and compact masses in the galaxy. These objects act as point-mass microlenses, introducing small-scale perturbative deflections. Though the resultant microimages of a microlensed quasar are too spatially clustered to be resolved, their combined magnification is observable. And, as microlenses move around in their respective lens planes, and the lens and source planes shift relative to each other, the brightness of the images varies over time \citep{chang_flux_1979, young_q0957561_1981, paczynski_gravitational_1986}. These temporal fluctuations can bias our inference of the cosmological parameters derived from the lightcurves \citep{liao_strong_2015, liao_impact_2021}.

To understand the effect microlensing would have on the observed lightcurves of the quasar lensed by a DSPL, we can write the deflection of an individual light ray as
\begin{equation}
    \mathbf{x}_s=\mathcal{M}\mathbf{x}
    -\sum_i\frac{m_i}{r_i^2}
    (\mathbf{x}-\mathbf{x}_i), \qquad
    r_i^2 = |\mathbf{x}-\mathbf{x}_i|^2,
    \label{eqn:lenseq}
\end{equation}
where
\begin{equation}
    \mathcal{M}= 
    \begin{pmatrix}
        1-\kappa_s-\gamma_1&-\gamma_2\\-\gamma_2&1-\kappa_s+\gamma_1
    \end{pmatrix}.
    \label{eqn:geo_matrix}
\end{equation}
The lens equation tells us the position of the source, $\mathbf{x}_s$, based on the observation of an image at $\mathbf{x}$. The first term accounts for the scaling due to a smooth matter distribution with convergence $\kappa_s$ and shear $(\gamma_1, \gamma_2)$. Note that $\mathcal{M}$ is mathematically equivalent to Equation \ref{eq:jacobian}, evaluated on a lens with respect to a source on the plane directly behind. However, the formulation in Equation~\ref{eqn:geo_matrix} does not assume that all matter is smoothly distributed; instead, the smooth and compact masses are treated separately, with the convergence contribution from compact matter accounted for by the sum in the second term. The summation over $i$ computes for the mass, $m_i$, and the physical distance to every microlens located at $\mathbf{x}_i$. We can calculate the shear strength, $\gamma=\sqrt{\gamma_{1}^{2}+\gamma_{2}^{2}}$, and angle, $\varphi_{\gamma}=\text{tan}^{-1}(\gamma_{2}/\gamma_{1})/2$, and rotate our coordinate system by $\varphi_{\gamma}$, such that the shear is aligned to the $x$-axis, and the off-diagonal terms in Equation~\ref{eqn:geo_matrix} are zero. The scaling matrix can then be rewritten:
\begin{equation}
    \mathcal{M} \mapsto\mathcal{M_\mathrm{aligned}}=
    \begin{pmatrix}
        1-\kappa_s+\gamma&0\\0&1-\kappa_s-\gamma
    \end{pmatrix}.
    \label{eqn:geo_matrix_align}
\end{equation}
The convergence term $\kappa_s$ in the scaling matrix depends on the critical surface density, which takes into account the angular diameter distances between the lens and source planes (Equation~\ref{eqn:kappa}--\ref{eqn:critdens}). Therefore, the lens equation is informed by cosmology, and constraining gravitational lensing will allow for constraints on cosmological parameters.

Equation~\ref{eqn:lenseq} is a one-to-many mapping, where a single source is microlensed into at least $N+1$ images, $N$ being the number of microlenses. This effect is recursive when considering lensing by consecutive planes; after lensing by the first plane, each resulting image acts as a new source for the subsequent plane. To retrieve the images, and hence the magnification of a microlensed source, we must first construct a physical model of the system to determine the relevant convergences and shears at each image position. We then take a numerical approach to solving this equation.

\section{Ray shooting methodology}
\label{sec:method}

Existing ray shooting microlensing codes consider single lens configurations \citep{wambsganss_microlensing_1990, vernados_new_2013, zheng_improved_2022, weisenbach_rootin_2025}, in which rays are generated on the lens plane, and traced to the source plane to obtain magnification maps, lightcurves and statistical properties for a given set of lens parameters. These approaches typically rely on vectorised implementations to efficiently evaluate a large number of rays.

We develop a new ray-shooting framework, written in \texttt{JAX}, leveraging Graphics Processing Unit (GPU) acceleration, and designed specifically for double-source-plane lensing. In this approach, rays are sequentially lensed at multiple planes (the lens plane and the s1 plane) before reaching the final s2 plane, requiring a consistent treatment of multi-plane deflections. The macrolens parameters and the numerical implementation of this new code are described in this section.

\subsection{Macrolensing Parameters}
\label{subsec:macromodel}

\begin{figure*}
    \centering
    \begin{subfigure}[b]{1\textwidth}
        \centering
        \includegraphics[width=\textwidth]{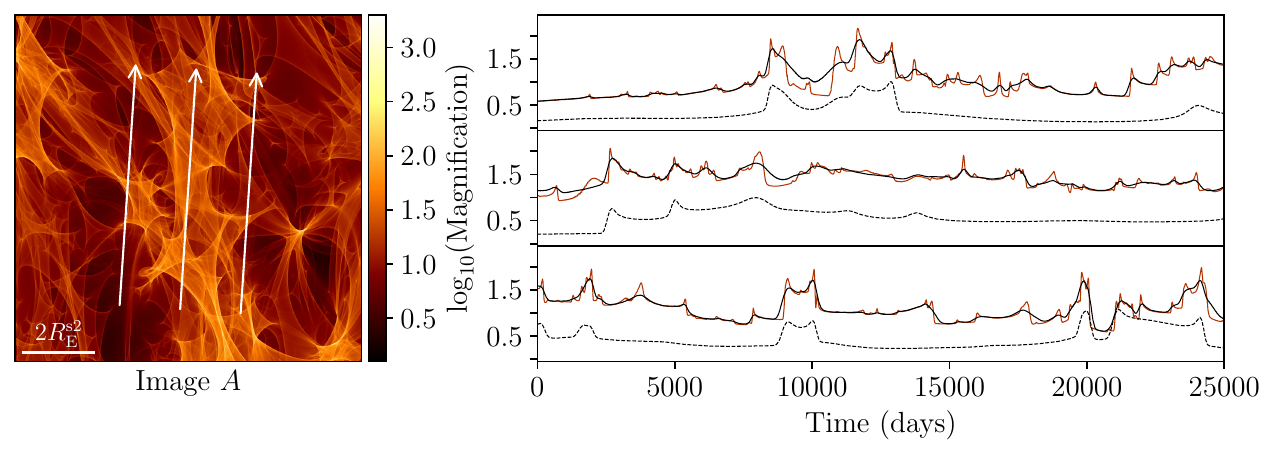}
    \end{subfigure}
    \begin{subfigure}[b]{1\textwidth}
        \centering
        \includegraphics[width=\textwidth]{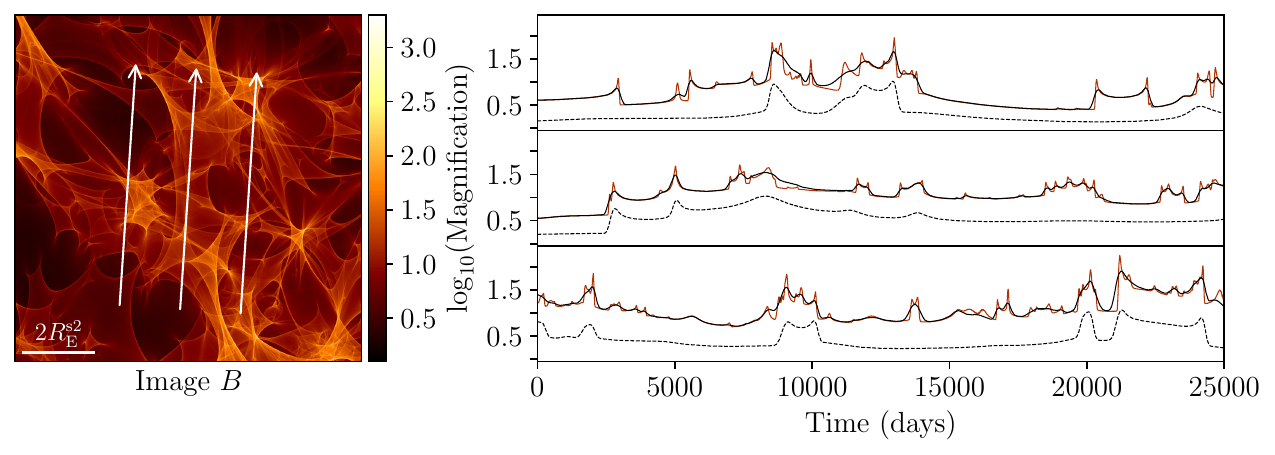}
    \end{subfigure}
    \begin{subfigure}[b]{1\textwidth}
        \centering
        \includegraphics[width=\textwidth]{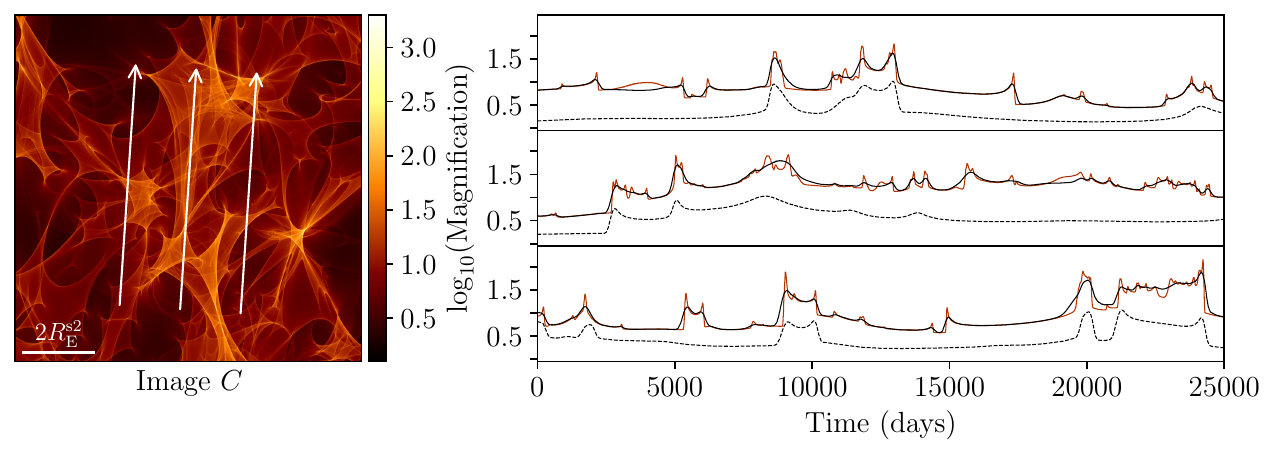}
    \end{subfigure}
    \caption{\textit{Left}: compound microlensing magnification maps for a quasar lensed by both the lens plane and s1 plane for images $A$--$C$. 
    Convergence and shear values at each plane are taken from Tables~\ref{tab:svi_median_kappa_gamma_lens} and \ref{tab:svi_median_kappa_gamma_s1}. The white lines on each map show three representative source tracks. \textit{Right}: magnification as a function of time for the three source tracks shown on the left. From top-to-bottom, the lightcurves correspond to the left-to-right order of the tracks on the maps. Dotted lines show the single-plane case, sampled from the maps shown in Figure~\ref{fig:s1maps}, and solid lines show the compound microlensing. The orange lightcurve is for a point source, where the black line is the lightcurve for the extended quasar source.
    }
    \label{fig:lightcurves}
\end{figure*}

We fit a macrolensing model to the HST ACS F814W observation of J1721 shown in Figure \ref{fig:J1721_data}. This is used to compute convergence, $\kappa$ and shear magnitude, $\gamma$ and angle, $\varphi$, at each of the images of the quasar as their rays pass through both lensing planes (see Appendix~\ref{app:macromodel} for details). In Table \ref{tab:svi_median_kappa_gamma_combined} we provide the effective lensing properties, which capture the combined lensing contribution of the full DSPL system. Table \ref{tab:svi_median_kappa_gamma_lens} gives the contributions from the lens plane alone, evaluated at the observed lens-plane quasar image positions. Table \ref{tab:svi_median_kappa_gamma_s1} gives the contributions from the s1 plane alone, evaluated at the mean locations of the two clusters of s1-plane image positions predicted by our model.

\begin{table}
\caption{The effective magnification, convergence and shear magnitude and direction $[\mathrm{deg}]$, as observed at the six image-plane quasar locations, predicted from our median macromodel result.}
\centering
\begin{tabular}{l|cccccc}
    & $A$ & $B$ & $C$ & $D$ & $E$ & $F$ \\ \hline
$\mu_{\text{eff}}$ & -14.508 & 10.866 & 11.486 & -4.161 & -2.666 & 1.672  \\
$\kappa_{\text{eff}}$ & 0.718 & 0.455 & 0.650 & 1.048 & 0.713 & 0.360 \\
$\gamma_{\text{eff}}$ & 0.520 & 0.392 & 0.272 & 0.571 & 0.682 & 0.816 \\
$\varphi_{\text{eff}}$ & 21.006 & -32.163 & -67.144  & 28.623 & -70.372 & -54.093 \\
\end{tabular}
\label{tab:svi_median_kappa_gamma_combined}
\end{table}

\begin{table}
\caption{Magnification, convergence and shear magnitude and direction $[\mathrm{deg}]$ of the foreground lens plane only, at the six image plane quasar locations, predicted from our median macromodel result.}
\label{tab:svi_median_kappa_gamma_lens}
\centering
\begin{tabular}{l|cccccc}
    & $A$ & $B$ & $C$ & $D$ & $E$ & $F$ \\ \hline
$\mu_{\text{lens}}$             & -6.178 & 4.085 & 3.970 & -1.410 & 1.878 & -1.973 \\
$\kappa_{\text{lens}}$          & 0.451 & 0.357 & 0.349 & 0.638 & 0.241 & 0.653 \\
$\gamma_{\text{lens}}$          & 0.680 & 0.411 & 0.414 & 0.917 & 0.206 & 0.792 \\
$\varphi_{\text{lens}}$ & 22.192 & -26.694 & 79.242 & 28.011 & -72.041 & 86.183 \\
\end{tabular}
\end{table}

\begin{table}
\caption{Magnification, convergence and shear magnitude and direction $[\mathrm{deg}]$ of the s1 plane only, at the two s1-plane quasar locations, predicted from our median macromodel result.}
\centering
\begin{tabular}{l|cc}
    & $ABCD$ & $EF$ \\ \hline
$\mu_{\text{s1}}$             & 2.687 & -1.055 \\
$\kappa_{\text{s1}}$          & 0.314 & 0.974 \\
$\gamma_{\text{s1}}$          & 0.314 & 0.974 \\
$\varphi_{\text{s1}}$ & -43.034 & -72.430 \\
\end{tabular}
\label{tab:svi_median_kappa_gamma_s1}
\end{table}

\subsection{Microlensing Code}
\label{subsec:rayshooting}

The natural length scale on the s1 and s2 planes is the Einstein radius of a $1M_\odot$ microlens on the plane immediately in front of it:
\begin{equation}
\label{eqn:einstein_radius}
    R_\mathrm{E}^\mathrm{s} = 
    \sqrt{
    \frac{4GM}{c^2}
    \frac{D(0,z_{s})D(z_{l},z_{s})}
    {D(0,z_{l})}
    },
\end{equation}
where $z_s$ is the redshift of the source plane (s1 or s2), and $z_l$ is the redshift of the lensing plane directly in front of it (lens or s1, respectively).

\begin{figure*}
    \centering
    \begin{subfigure}[b]{1\textwidth}
        \centering
        \includegraphics[width=\textwidth]{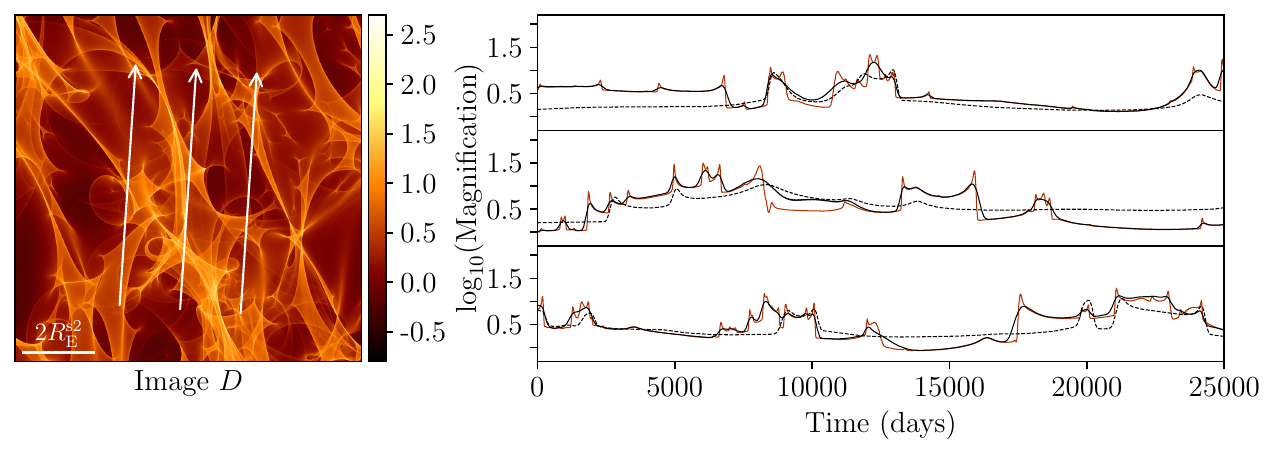}
    \end{subfigure}
    \begin{subfigure}[b]{1\textwidth}
        \centering
        \includegraphics[width=\textwidth]{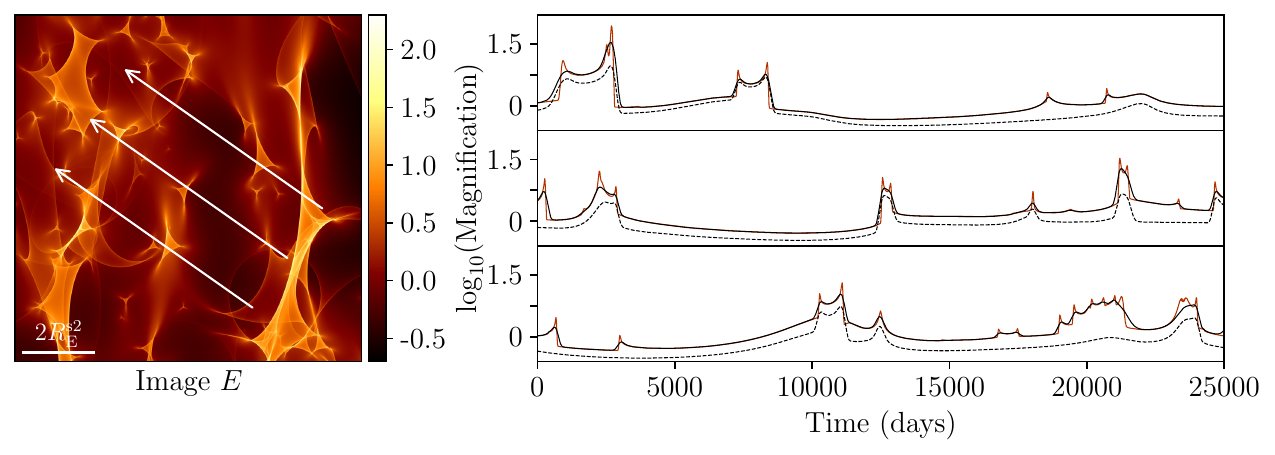}
    \end{subfigure}
    \begin{subfigure}[b]{1\textwidth}
        \centering
        \includegraphics[width=\textwidth]{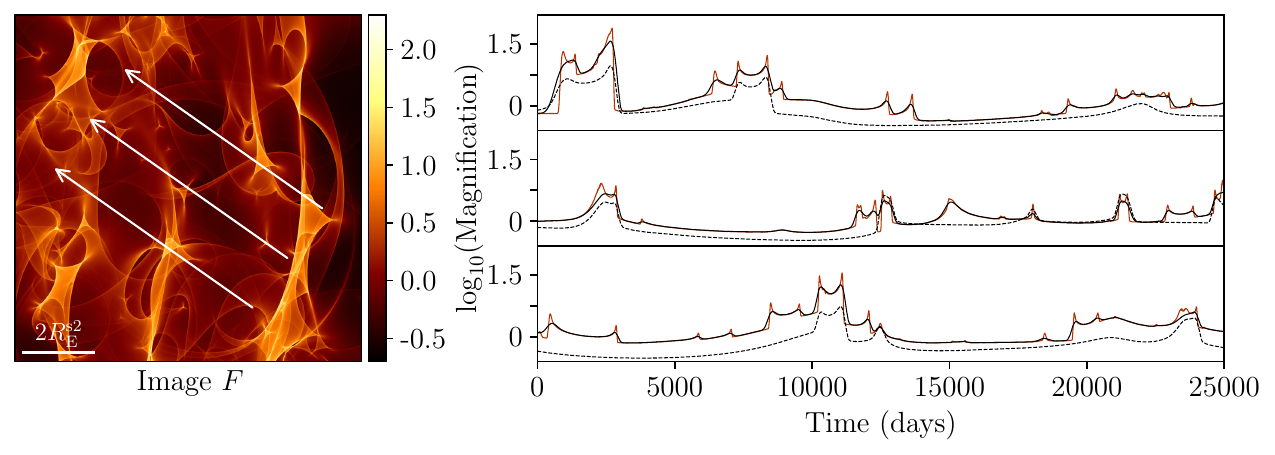}
    \end{subfigure}
    \caption{Likewise as in Figure~\ref{fig:lightcurves}, for images $D$--$F$. For images $E$ and $F$ the lightcurves correspond to the top-to-bottom order of the tracks on the maps.} \label{fig:lightcurves2}
\end{figure*}

In our s1 shear-aligned frame, we define the s2 plane as a square with side length $L_s=10R_\mathrm{E}$. The size of the s1 plane follows from the inverse of the geometric transformation matrix in Equation~\ref{eqn:geo_matrix_align}, such that the s1 plane is a rectangle with sides
\begin{equation}
    L_{x,y}=L_s|1-(\kappa_s+\kappa_*\pm\gamma)|^{-1}.
\end{equation}
We compute the shape of the lens plane similarly to the inverse of the transformation matrix in Equation~\ref{eqn:geo_matrix} applied to the s1 plane, transforming the lens plane into a parallelogram. The total area of the lens plane is scaled by a factor $\left|\det(\mathcal{M})^{-1}\right|$ relative to the s1 plane, which is equivalently the magnitude of the macromagnification of the foreground lens acting on s1.

Assuming that all of the convergence is due to $1M_\odot$, static microlenses, such that $\kappa_*=\kappa$ and $\kappa_s=0$, we generate a randomised uniform field of microlenses on each plane in a rectangular region containing that plane. The number of microlenses is determined by the surface density $\kappa$. While a more rigorous treatment would take a circular distribution of microlenses, this would require a significantly larger number of masses. For this work, the improvement is negligible relative to the added computational cost.

We generate a grid over the lens plane, within which we place a regular mesh of fictitious image positions (or ``rays''). The ray shooting region is slightly more extended than the calculated lens plane, accounting for edge effects, but it fits well within the field of microlenses. These rays are mapped to the source plane directly behind them via Equation~\ref{eqn:lenseq}. The final map is then obtained up to a global scale factor, which is the ratio of the density of rays on the lens plane and the size of a pixel on the source plane. 
 
The addition of a second source plane, and the successive expansion between the s2, s1 and lens planes substantially increases the microlens sampling region, and hence the number of microlenses for which the deflections are computed and summed. Unlike standard single plane ray shooting, which reduces to a direct summation over microlens deflections, this extended formalism means rays are mapped twice, and the deflections and summations must be computed twice, over a substantial number of lenses. As no existing implementations support this multi-plane configuration at the required scale, we develop a bespoke ray shooting framework. To make this tractable, we divide the rays into tiles, and we vectorise and evaluate both the microlens summation term and the full set of rays in a tile using \texttt{JAX}'s \texttt{vmap}. We stream and process the tiles in parallel, allowing batches of rays to be evaluated efficiently.

We make three maps for each image: one from the s1 to s2 plane, and one from the lens to the s1 plane. The third map is the magnification of a compoundly-microlensed source, and takes rays generated on the lens plane, maps them to the s1 plane, and then further maps these onto the s2 plane. Each of these maps contains the positions of at least $10^{10}$ rays, and is split into $10^4\times10^4$ pixels.

The quasar source is modelled as a Gaussian profile, adopting a physical radius of $10^{15}\mathrm{cm}$ \citep{paczynski_gravitational_1986, wambsganss_microlensing_2001}, corresponding to a standard deviation $\sigma_\mathrm{s}=0.0115R_\mathrm{E}$, or about 23 pixels. To simulate lightcurves of the quasar, we first convolve our magnification map with this Gaussian. The quasar is set in motion along three horizontal straight-line trajectories across the magnification map. Because each macroimage is expressed in a frame rotated to align with the s1 shear, this fixed trajectory is represented at a different angle in images $A$--$D$ in comparison to images $E$ and $F$. We choose a transverse velocity of $600\mathrm{km/s}\approx2\times10^{-4}\mathrm{R_{E}/day}$, the value typically used for generating microlensing lightcurves at such cosmological distances \citep[see.][]{kayser_astrophysical_1986, schneider_gravitational_1987}. Given the source's trajectory, we can sample the pixel value of the convolved map along the line to obtain the magnification of the source over time.

\section{Results}
\label{sec:results}

We present two sets of magnification maps: (i) the magnification map around the two images created when the quasar in the s2-plane is lensed by masses in the s1-plane, shown in Figure~\ref{fig:s1maps}, and (ii) the full compound microlensing magnification maps of the quasar lensed by the foreground lens after said precursive lensing by the s1-plane, shown in the left column of Figure \ref{fig:lightcurves} and \ref{fig:lightcurves2}. The right column of Figure \ref{fig:lightcurves} and \ref{fig:lightcurves2} additionally shows three example lightcurves of the quasar when microlensed by s1 alone (dotted), and as a consequence of compound microlensing by both the foreground lens and s1 plane (solid). These compound microlensing lightcurves are shown for both a point source (orange) and our extended quasar source (black).

\begin{figure}
    \centering
    \includegraphics[width=\linewidth]{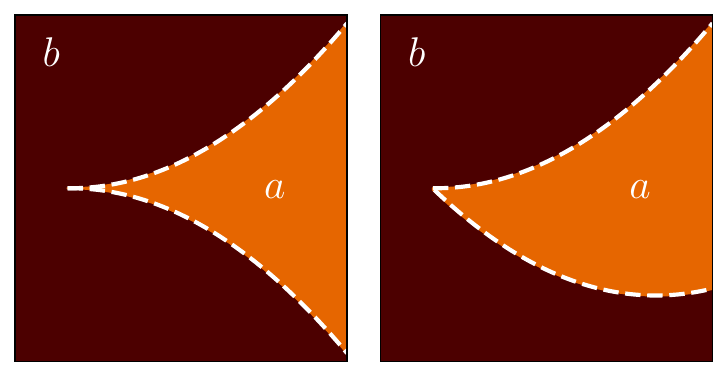}
    \caption{Illustration of convexity violation seen in compound microlensing. Here, the inside of a caustic is labelled as region $a$, and the outside is region $b$. \textit{Left:} caustic of the first kind, where both folds are convex relative to region $a$. \textit{Right:} caustic of the second kind, where one fold is concave relative to region $a$.
    }\label{fig:conv_violation}
\end{figure}

\begin{figure*}
    \centering
    \includegraphics[width=\textwidth]{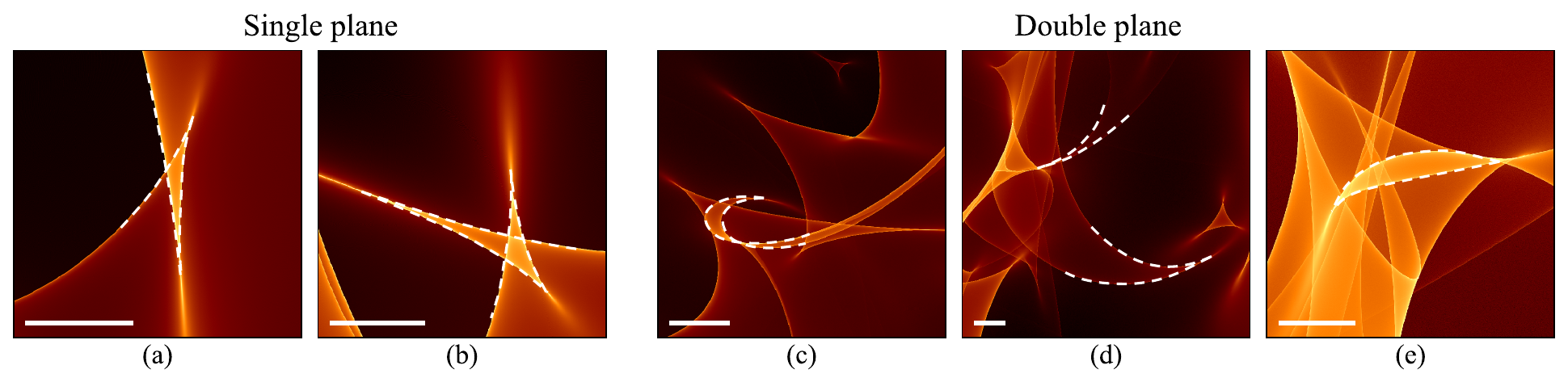}
    \caption{Examples of caustic structures in the microlensing magnification maps, relevant features are outlined in a white dashed line. (a) A swallowtail caustic in map $ABCD$. (b) A butterfly caustic in map $ABCD$. (c)–(e) Convexity-violating caustics in maps $C$, $E$, and $F$, respectively. Panel (e) shows a lip caustic, which is not permitted in single-plane lensing \citep{petters_caustics_1995}. The white bar in each panel corresponds to a length of $0.25\,R_\mathrm{E}^\mathrm{s2}$. 
    }\label{fig:pretty_arcs}
\end{figure*}

For all quasar images, the compound microlensing maps in Figure~\ref{fig:lightcurves} and \ref{fig:lightcurves2} retain similar caustic structure as observed in the corresponding s1-plane maps in Figure~\ref{fig:s1maps}, but with the addition of a secondary caustic network from the extra foreground lens plane. We identify convexity violations in our compound microlensing maps, a clear signature of multi-plane lensing, illustrated in Figure~\ref{fig:conv_violation}.  In a single plane magnification map, one can only observe caustics of the ``first kind''. Here, from the perspective of the inside of the caustic (region a), the two folds meeting at a cusp exhibit uniform convex curvature (left panel). In contrast, double-plane microlensing may produce ``caustics of the second kind'' \citep{petters_caustics_1995}, in which one of the two folds meeting at a cusp becomes locally concave relative to the inside of the caustic (right panel). Such convexity violations are features exclusive to our compound magnification maps, with representative examples, shown in the right three panels of Figure~\ref{fig:pretty_arcs}; Examples can also be found throughout our magnification maps in Figures \ref{fig:lightcurves} and \ref{fig:lightcurves2}.

In contrast, the single-plane maps exhibit only the standard caustic metamorphoses expected from single-plane lensing, including swallowtail and butterfly structures in the left two panels of Figure~\ref{fig:pretty_arcs}. The recursive deflections across two planes introduce new structures that are not permitted in single-plane scenarios \citep{petters_caustics_1995}, such as a lip caustic shown in the right-most panel in Figure~\ref{fig:pretty_arcs}. As our analysis is based on static magnification maps, rather than time-evolving caustic crossings, we do not track individual caustic metamorphoses; instead, these features appear as distinct local morphologies within the caustic network.

Since there is no intrinsic alignment between the shear in the foreground lens and s1 planes, the compound microlensing magnification maps include significant shearing in multiple directions, not just perpendicular to the motion of the quasar, as in Figure \ref{fig:s1maps}. Image $E$, coincidentally, has very similar shear angles on both lens planes. Still, the convergence of the lens plane is low at this location, so it is difficult to compare the resultant map to those of the other images with unaligned shears. It is noteworthy, however, that in theory an image can be produced with similar observed effective magnifications even if the shear alignment gives a very different caustic structure.

The compound microlensing lightcurves broadly follow the same trend as the singly-microlensed lightcurves, where the caustic structure on the magnification map is retained; some of its features are emphasised after the second microlensing event, which also adds an abundance of smaller-scale peaks across the lightcurve. As a result, some of the features from the microlensing of s1 are less pronounced when intermingled with the additional microlensing of the foreground lens. For both images $D$ and $F$, the compound microlensing lightcurves are at some points less magnified than the corresponding single-lens case. We also note that these two maps tend to produce more convexity-violating cusps. Since we are testing a small parameter space, it is currently unclear whether this effect is simply a coincidence or due to the fact that these are the two zig-zag images. These points will be subject to further study.

\section{Conclusions}
\label{sec:discussion}

Strong gravitational lensing is a powerful probe of cosmological parameters and galaxy evolution: compound lenses provide robust constraints on foreground lens potentials, or yield $H_{0}$-invariant constraints on the equation of state for dark energy; time-delay lenses have contributed significantly to late-Universe measurements of $H_{0}$; microlensing can probe the accretion disk sizes of high-redshift active galactic nuclei. In this work, we consider the occurrence of microlensing on a compoundly-lensed time-varying quasar, and perform, to our knowledge, the first practical demonstration of compound microlensing: the microlensing performed by multiple planes of compact masses in succession. The recently discovered ``Einstein zig-zag'', J1721+8842, provides a fiducial testbed of two strong (multiply-imaging) lenses along the line of sight to a quasar, which appears sextuply-imaged.

We fit a basic macromodel of J1721 with six image magnifications broadly consistent with the comprehensive lens modelling study of \citet{schmidt_TDCOSMO_2025}. The convergence and shear at two locations where images of the quasar form on the intermediate lens plane, and six locations where the images form at the foreground lens plane, are measured. Using these values, we create physically viable microlensing simulations of six images lensed by two sets of planes.

A GPU-accelerated inverse ray shooting pipeline written in \texttt{JAX} was used to simulate compound microlensing for the first time. The computational cost of producing each map scales with both the number of microlenses and the number of rays. We find that the sequential lensing in the compound case typically leads to larger effective lens plane regions than in single plane models, as the size of each plane is dictated by magnification. However, since microlenses must be distributed across a rectangular region enclosing the lens plane (plus an additional border width for numerical stability), even images with relatively low magnification can require a large number of microlenses when the shear magnitude is high, as the bounding rectangle of the sheared region can be significantly extended. Since we are simulating quasar microlensing, the source-plane map requires a fine pixel scale to resolve the compact source. This in turn demands a large number of rays, as finer pixels receive fewer rays on average and are therefore more susceptible to Poisson noise. Future work could reduce the per-ray computation cost via alternative approaches (e.g. inverse polygon mapping; \citet{weisenbach_rootin_2025}, tree-like interpolation; \citet{ zheng_improved_2022}).

While \citet{petters_caustics_1995, petters_singularity_2001} provide a theoretical multi-plane classification for isolated point-mass lenses, we practically demonstrate that these features persist in realistic, high-density microlensing environments. Using microlensing simulations of a realistic DSPL, we concur with theoretical predictions that swallowtail and butterfly caustic metamorphoses are ubiquitous across single- \textit{and} multi-plane microcaustic maps, but that there are additionally convexity-violating caustic structures and more exotic metamorphoses such as lips which only appear as a result of compound microlensing. Correspondingly, our simulated lightcurves display two signatures of a second lens plane: an overall amplification of the peaks induced by the first lens plane, as well as an additional population of peaks produced by new caustic-crossing events.

The statistical interpretation of our magnification map structures and lightcurves is non-trivial, but the qualitative analysis performed in this work makes it obvious that observed differences between single- and multi-plane microlensing are significant. It is currently unclear how much convexity violation will be suppressed under a more realistic assumption that the convergence is formed of a smooth distribution of matter as well as compact masses. It is also not clear what features we expect from magnification maps with different shear alignments between the lens and s1 planes, as they may act non-linearly to amplify or suppress local magnifications. 

Clearly, there is an extensive parameter space yet to be explored. For instance, typical DSPL systems have late-type star-forming galaxies on their intermediate planes, unlike in J1721. These have significantly different morphological properties from the foreground lenses, which are typically early-type elliptical galaxies. This motivates exploring different initial mass functions to generate the stellar population of microlenses on each plane, as well as different contributions from the smooth mass component, which we have not modelled in this work. Existing NIRSpec data from JWST can give insight into the velocity dispersion of the stellar population. This would allow us to simulate a more realistic situation in which we can demonstrate how the relative motion of microlenses will affect our statistics. 

To close, we recall that compoundly-lensed objects are promising cosmological probes, and quasar microlensing is an imperative tool to understanding the physical scales of active galactic nuclei. To fully unlock the potential of a statistically inevitable sample of compoundly lensed quasars, we will need a thorough understanding of the microlensing effect on the measured flux of the multiple images. In this work, we have implemented the first recursive ray-shooting to practically demonstrate that compound microlensing produces new and more intricate microcaustic structures than those found in single-plane microlensing. A full statistical characterisation of compound microlensing will require extending this analysis across a larger parameter space, requiring substantial additional computation.

\section*{Acknowledgments}
We thank Smrithi Gireesh Babu and Oliver Oayda for their helpful comments on this manuscript. NS acknowledges support from the University of Sydney through the Faculty of Science Research Stipend Scholarship. HQ gratefully acknowledges the support of the International Research Training Program Scholarship.
DJB, GFL and KG acknowledge the support of the Australian Research Council Discovery Project DP230101775. DJB and GFL also thank the Center for Astrophysics at Harvard University for hosting the ``Strong Lensing in the Next Decade'' 2025 workshop, where the seeds of this project were sown. 

This research was undertaken with the assistance of resources from the National Computational Infrastructure (NCI Australia), an NCRIS enabled capability supported by the Australian Government and the Sydney Informatics Hub, a Core Research Facility of the University of Sydney.

\section*{Data Availability}

The data generated for this paper, and the code used to generate the data, will be made available on reasonable request to the corresponding author.



\bibliographystyle{mnras}
\bibliography{bibliography} 



\appendix
\section{Macromodel Details}
\label{app:macromodel}

\begin{figure*}
    \centering
    \includegraphics[width=0.99\linewidth]{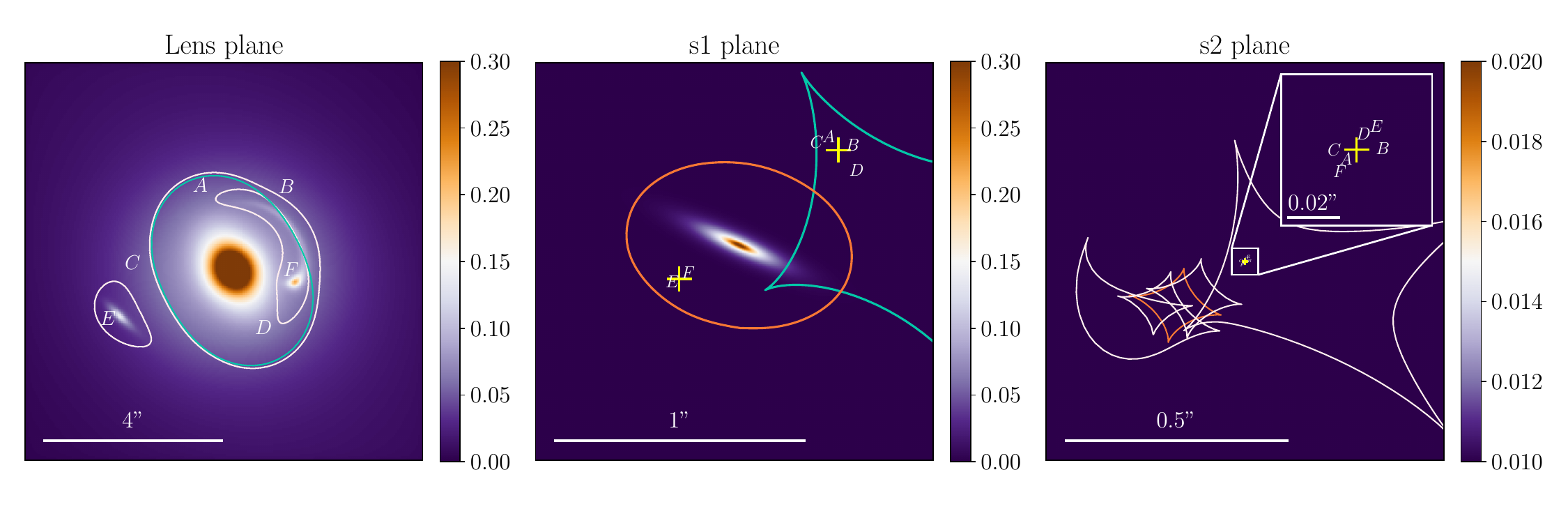}
    \caption{The median SVI result of our macromodel for J1721. In the left panel, the surface brightness of the best fit lens light and arcs are shown with the six quasar positions ($A$--$F$) from the data overlaid as in Figure~\ref{fig:J1721_data}. In the middle panel, we show the median reconstruction of the s1-plane with the surface brightness of the source and quasar locations $A$--$F$ corresponding to those of the image plane; note that they form two images, the average location of images $A$--$D$ and $E$--$F$ are denoted by yellow crosses. The right panel shows the s2-plane locations of images $A$--$F$ and their average location at the yellow cross, which are also shown in the zoomed inset panel. White contours show the critical curve (left) and caustic (right) for the full multi-plane model; the cyan critical curve (left) and caustic (middle) are for the foreground lens acting on s1 only; the orange critical curve (middle) and caustic (right) are for the lens on the s1 plane acting on s2 only. For computational reasons, and to avoid visual clutter, we have only shown the tangential critical curves and corresponding caustics; the radial critical curves and corresponding caustics are omitted.}
    \label{fig:macromodel}
\end{figure*}
To obtain the relevant parameters to simulate compound microlensing in a DSPL, we need a macromodel of our fiducial system. Our macromodel of J1721 successfully focuses the six images of the quasar to within $\sim0.02^{\prime\prime}$ in the s2-plane. At s1, we obtain two clusters of images of the quasar, which are not as well focused, but their two mean locations provide sufficiently realistic deflector mass properties for this plane. For a dedicated lens modelling analysis of J1721, we direct the reader to \citet{schmidt_TDCOSMO_2025}. Given the simplistic assumptions of our model, we do not anticipate that our convergence, shear or magnification estimates will agree with this work, though we do find that our effective quantities in Table \ref{tab:svi_median_kappa_gamma_combined} are broadly consistent.

We assume that the mass of the foreground lens can be described by an Elliptical Power Law \citep[EPL;][]{tessore_elliptical_2015} mass profile, whose convergence is given by:
\begin{equation}
\label{eqn:kappa_EPL}
    \kappa_{\text{EPL}}(r)=\frac{3-\gamma}{2}\left(\frac{\vartheta_{E}}{r}\right)^{\gamma-1},
\end{equation}
where $\gamma$ is the logarithmic slope, $\vartheta_{E}$ is the Einstein radius, and $r=\sqrt{q^{2}x^{2}+y^{2}}$ is the elliptical radius, defined using 2D cartesian coordinates $(x, y)$ and the minor-to-major axis ratio of the profile, $q$. We allow the centre coordinate of the lens, $(x_0, y_0)$, as well as its position angle, $\varphi$, to also vary freely.

We additionally assume that the foreground lens is embedded within a sheared field, parameterised by external shear strength, $\gamma^{\text{ext}}$, and position angle, $\varphi^{\text{ext}}$. To avoid bimodality when sampling over angles, we re-parameterise $q$ and $\varphi$ as $(e_{1}, e_{2})$ and $\gamma^{\text{ext}}$ and $\varphi^{\text{ext}}$ as $(\gamma^{\text{ext}}_{1}, \gamma^{\text{ext}}_{2})$:
\begin{equation}
    (e_{1}, e_{2}) = \frac{1-q}{1+q}(\sin{2\varphi}, \cos{2\varphi});
\end{equation}
\begin{equation}
    (\gamma^{\text{ext}}_{1}, \gamma^{\text{ext}}_{2}) = \gamma^{\text{ext}}(\sin{2\varphi^{\text{ext}}}, \cos{2\varphi^{\text{ext}}}).
\end{equation}

We model the mass distribution of s1 as the $\gamma=2$ case of Equation~\ref{eqn:kappa_EPL}, also known as the singular isothermal ellipsoid (SIE) profile. 

Our constraints on this model are the appearance of the lens and the lensed pair of images of s1, as well as the locations of the brightest pixel within each quasar image $A$ through $F$. The brightnesses of the quasar images, which are susceptible to microlensing magnification, and accurate modelling of the PSF are ignored. 

We model the light distribution of the foreground lens using a multi-Gaussian profile as in \citet{he_unveiling_2024}. We use 20 components, which are concentric to one centroid coordinate but are allowed to vary in scale, position angle and axis ratio. We assume that the light distribution of s1 is described by an elliptical Sérsic profile:
\begin{equation}
    s(r) = s_{e}\exp{\left[-b_{n}\left(\left(\frac{r}{r_{e}}\right)^{1/n}-1\right)\right]},
\end{equation}
where $n$ is the Sérsic index and $r_{e}$ is the effective radius, or the half-light radius, defined as the radius that encloses the effective surface brightness, $s_{e}$, which is half the total surface brightness of the profile. The coefficient $b_{n}$ is approximated as $b_{n}=1.9992n-0.3271$.

We add a minimisation term to our likelihood which penalises unfocused quasar images. This term assumes that the mean distance, $\langle d_{s2}\rangle$, between the s2-plane positions of the six quasar images follows an exponential distribution, $\langle d_{s2}\rangle \sim \exp(\lambda)$, where $\lambda$ is the rate parameter. Through experimentation, we set $\lambda=500$, which we find effectively penalises lens model solutions that are poorly focused at the s2 plane.

This model is fit using the $\texttt{numpyro}$ \citep{phan_composable_2019} implementation of stochastic variational inference \citep[SVI;][]{hoffman_stochastic_2013} within a multi-plane extension of the strong lens modelling framework \texttt{herculens}\footnote{\url{https://github.com/Herculens/herculens}} \citep{galan_herculens_2022}, written in \texttt{JAX}. Within the SVI, we use \texttt{AdaBelief} to optimise a low-rank multivariate normal distribution
to the true posterior.

The median of this distribution is used to calculate the convergences and shears we require. In Figure~\ref{fig:macromodel}, we show the critical curves and caustics of each plane, and the resulting quasar positions, for this median model.


\bsp	
\label{lastpage}
\end{document}